\pdfoutput=1
\documentclass[12pt,draftclsnofoot,onecolumn]{IEEEtran}
\usepackage{amssymb}
\usepackage{amsmath}
\usepackage{amsfonts}
\usepackage{graphicx}
\usepackage{subfigure}
\usepackage{cite}
\usepackage{enumerate}
\usepackage{}
\usepackage{url}
\usepackage[ruled]{algorithm2e}
\usepackage{algorithmicx}
\usepackage{mathtools}
\usepackage{makecell}
\usepackage{bm}
\usepackage{color}
\allowdisplaybreaks[4]

\newif\ifthickvlines
\newcolumntype{|}{!{%
  \ifthickvlines
    \vrule width2\arrayrulewidth
  \else
    \vline
  \fi}}
\newcommand\thickvlines{\noalign{\global\thickvlinestrue}}

\newcommand{\blue}{\textcolor[rgb]{0,0,0}}

\begin{document}

\newtheorem{definition}{\bf Definition}	
	
\title{\huge Reinforcement Learning for a Cellular Internet of UAVs: Protocol Design, Trajectory Control, and Resource Management}
\author{
\IEEEauthorblockN{
\normalsize{Jingzhi Hu}\IEEEauthorrefmark{1},
\normalsize{Hongliang Zhang}\IEEEauthorrefmark{1}\IEEEauthorrefmark{2},
\normalsize{Lingyang Song}\IEEEauthorrefmark{1}, 
\normalsize{Zhu Han}\IEEEauthorrefmark{2},
and \normalsize{H. Vincent Poor}\IEEEauthorrefmark{3}\\}
\IEEEauthorblockA{\small
\IEEEauthorrefmark{1}Department of Electronics, Peking University, Beijing, China \\
\small\IEEEauthorrefmark{2}Department of Electrical and Computer Engineering, University of Houston, Houston, TX, USA\\
\small\IEEEauthorrefmark{3}Department of Electrical Engineering, Princeton University, Princeton, NJ, USA\\}
}
 
\maketitle
%\vspace{-0.2cm}

\begin{abstract}
Unmanned aerial vehicles (UAVs) can be powerful Internet-of-Things components to execute sensing tasks over the next-generation cellular networks, which are generally referred to as the cellular Internet of UAVs.
However, due to the high mobility of UAVs and the shadowing in the air-to-ground channels, UAVs operate in an environment with dynamics and uncertainties.
Therefore, UAVs need to improve the quality of service~(QoS) of sensing and communication without complete information, which makes reinforcement learning suitable to be employed in the cellular Internet of UAVs.
In this article, we propose a distributed sense-and-send protocol to coordinate the UAVs for sensing and transmission.
\blue{Then, we apply reinforcement learning in the cellular Internet of UAVs to solve key problems such as trajectory control and resource management.}
Finally, we point out several potential future research directions.
\end{abstract}

\newpage
%%%%%%%%%%%
\section{Introduction}
%%%%%%%%%%%
\begin{figure}[t!]
	\centering
\includegraphics[width=0.8\linewidth]{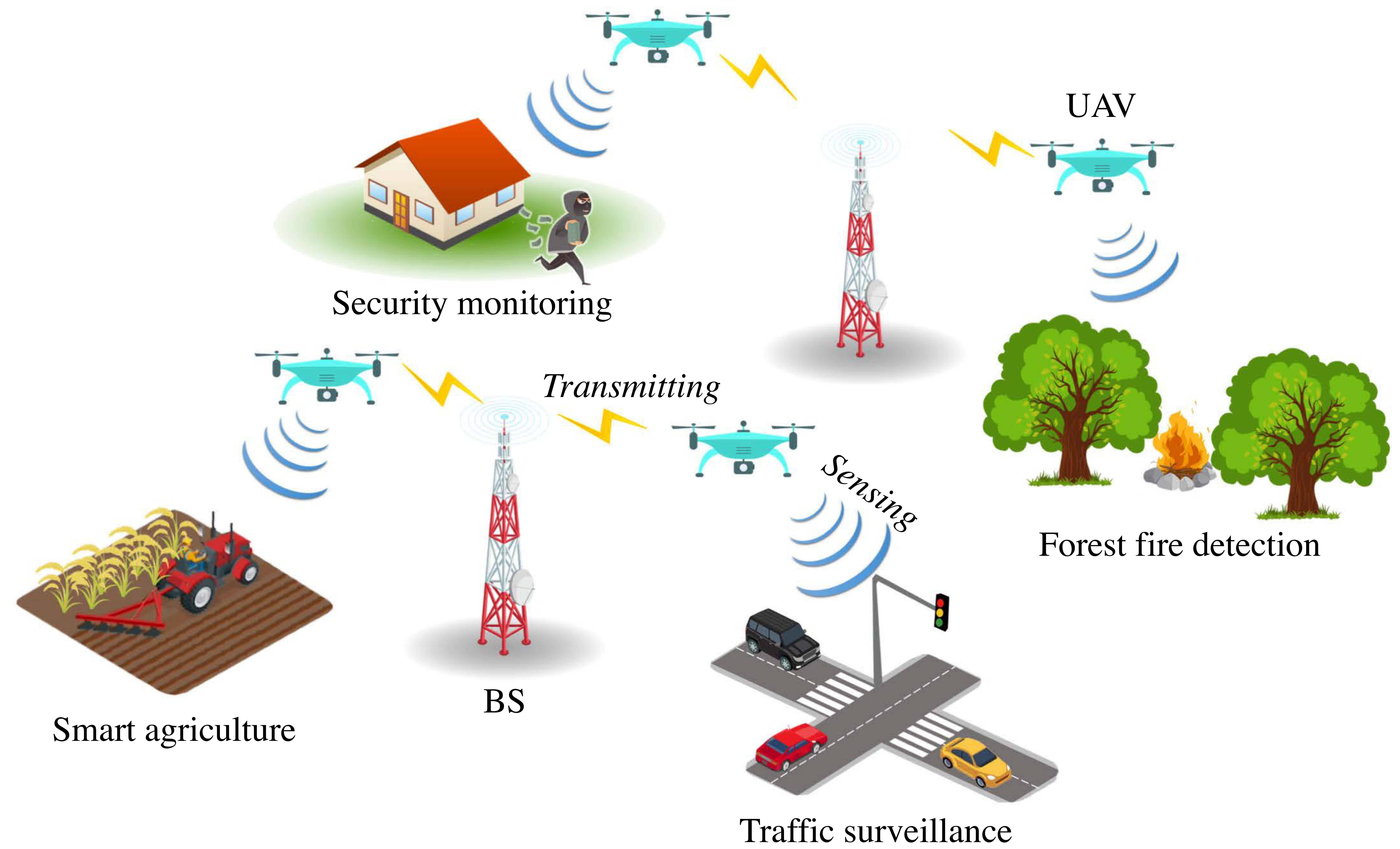}
%\vspace{-1em}
	\caption{Cellular Internet of UAVs for various kinds of sensing tasks.} \label{network1}
	\vspace{-1em}
\end{figure}
The emerging unmanned aerial vehicles (UAVs) have been playing an increasing role in the military, public, and civil applications~\cite{SER-2016}.
Specially, exploiting UAVs as Internet of Things~(IoT) devices to execute sensing tasks has been of particular interests due to its advantages of on-demand flexible deployment, large service coverage, and capability to hover at high altitude~\cite{Wang2017Taking}.
Such sensing tasks include a wide range of critical daily applications, e.g., smart agriculture, security monitoring, forest fire detection, and traffic surveillance, as illustrated in~Fig.~\ref{network1}.
To realize the above visions, it is envisaged and studied by the Third Generation Partnership Project~(3GPP) that the cellular networks are necessary for UAVs to execute sensing tasks, which we refer to as the cellular Internet of UAVs~\cite{Zhang2019Cooperation}. 

%ä»ç»äºæ äººæºè¿è¡ä»»å¡çæ­¥éª¤
In the cellular Internet of UAVs, the UAVs sense the targets of the tasks and then transmit the sensory data to the base stations~(BSs) immediately.
%æä»¥æç¥åä¼ è¾æ¯coupledå¨ä¸èµ·ç
Therefore, the sensing and transmission of UAVs are coupled~\cite{Zhang2019Cellular}.
%ç¯å¢çå¨ææ§åæªç¥æ§
Besides, due to the high mobility of UAVs and the shadowing in the air-to-ground channels, UAVs operate in an environment with dynamics and uncertainties~\cite{Mozaffari2018ATutorial}.
%æä»¥å¢ï¼è¿å¯¼è´äºä»ä¹é®é¢ï¼
Therefore, UAVs need to improve the quality of service~(QoS) of both sensing and transmission without complete information,
%è¿ä¸ªé®é¢ä½¿å¾coordinationå¾é¾
which makes the coordination of multiple UAVs to execute sensing tasks a challenging problem.

In this article, we introduce the reinforcement learning approaches and their applications in the cellular Internet of UAVs.
Since reinforcement learning can enable UAVs to improve their policies for their objectives without a priori knowledge of the environment, it is suitable to address the key problems in the cellular Internet of UAVs~\cite{Sutton2018Reinforcement}, such as trajectory control and resource management.
\blue{Specifically, we focus on the following three essential parts in the cellular Internet of UAVs.
\begin{itemize}
\item \textbf{Protocol Design}: We present a distributed sense-and-send protocol to coordinate the UAVs in sensing and transmission.
\item \textbf{Trajectory Control}: We discuss the dynamic trajectory control problem of the UAVs and propose an enhanced multi-UAV Q-learning algorithm for this problem. 
\item \textbf{Resource Management}: We introduce different types of reinforcement learning approaches and their applications for resource management problems, including user association, power management, and subchannel allocation.
\end{itemize}
}

To address the trajectory control and resource management, we discuss the possible implementations of reinforcement learning approaches in the cellular Internet of UAVs:
	1) applying multi-armed bandit learning to solve the user association problem;~2) utilizing Q-learning to solve the trajectory control problem;~3) using actor-critic learning to solve the power management problem; and~4) applying deep reinforcement learning to solve the subchannel allocation problem.

The rest of article is organized as follows. 
In Section~\ref{sec: Scenario}, we provide an overview of the cellular Internet of UAVs, and demonstrate the sense-and-send protocol. 
In Section \ref{sec: RL and app.}, we discuss the reinforcement learning approaches, including the basics and applications in the cellular Internet of UAVs. 
We then in Section~\ref{sec: Q-example} elaborate on how to apply Q-learning to solve the UAV trajectory control problem.
In Section~\ref{sec: Conclusion}, we draw the conclusions and point out several future research directions.

%%%%%%%%%%%%%%%%%%%%%%%%%
\section{Overview of the Cellular Internet of UAVs}
%%%%%%%%%%%%%%%%%%%%%%%%%
\label{sec: Scenario}
In this section, we first introduce the cellular Internet of UAVs.
Then, to coordinate of multiple UAVs to execute sensing tasks, we propose a distributed sense-and-send protocol.
%===========================
\subsection{Cellular Internet of UAVs}
%===========================
 \begin{figure}[!t]
	\centering
	\includegraphics[width=0.767\linewidth]{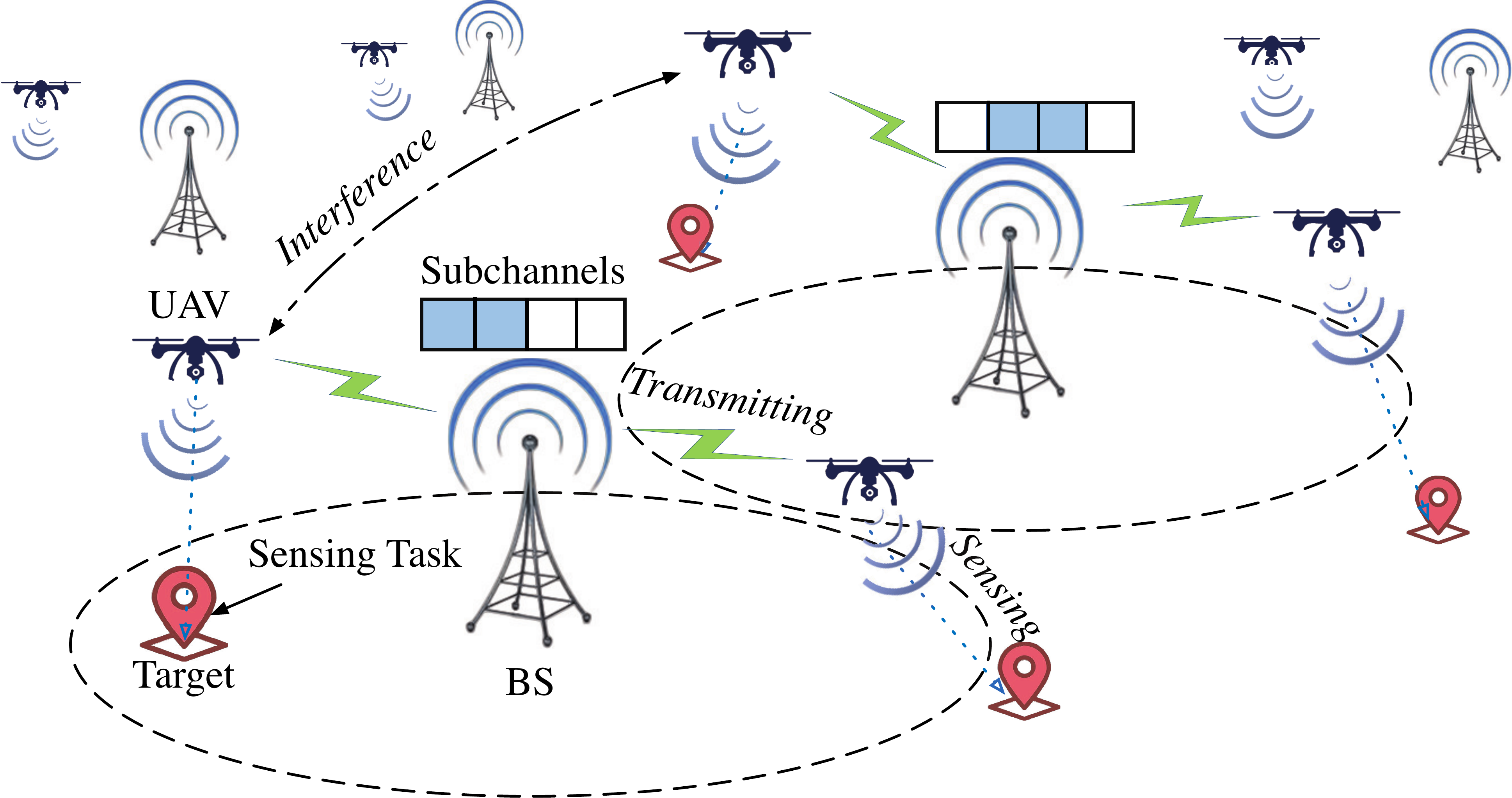}
	\vspace{-0.5em}
	\caption{System model of the cellular Internet of UAVs.} \label{fig: sys mod}
		\vspace{-1em}
\end{figure}
%General
%TODO add more words
As shown in Fig.~\ref{fig: sys mod}, in a cellular Internet of UAVs, multiple UAVs execute a set of sensing tasks.
Each task has one target to be sensed, and the targets are at different locations.
\blue{The tasks are pre-assigned to the UAVs, and the UAVs sense the targets of their tasks continuously and transmit the results to the BSs.}
Therefore, the UAVs execute the tasks through two steps, i.e. \emph{UAV sensing} and \emph{UAV transmission}:

%UAV Sensing 
\begin{itemize}
	%..........................................
	\item \textbf{UAV Sensing}: 
	%..........................................
	Each UAV is equipped with an on-board sensor.
	Due to the limited sensing capability of the sensor, the sensing is not always successful\footnote{
	%footnote begin
	%æ¾ä¸ä¸æ¾å°çå ç¯ç¨æ¥åç­sensing æ¨¡åçæç« ï¼çä¸ä¸éé¢æä¹å®ä¹successful sensingçã
	For example, the sensing is considered to be successful when the sensor successfully detects an event~(e.g., a traffic jam at a crossroad), or correctly measures the condition of a target~(e.g., the air quality at a certain location).
	%footnote end
	}.
	If the sensing is successful, the sensory data collected by the UAV is referred to be \emph{valid}; otherwise, it is referred to be \emph{invalid}.
	In general, the probability of successful sensing is negatively related to the distance between the sensor and the target~\cite{VI-2017}.
	%.................................................
	\item \textbf{UAV Transmission}:
	%.................................................
	%UAV Transmission
	Each UAV is associated with one BS and uses the uplink subchannels allocated by the BS to transmit the sensory data\footnote{\blue{In the cellular Internet of UAVs, since the UAVs are IoT devices, the uplink transmission for sensory data dominates. Therefore, in this paper, we focus on the uplink transmission of the UAVs.}}.
	Each BS owns a limited number of subchannels to support UAV transmission.
	The frequency bands used by different BSs can be overlapped or orthogonal, determined by the deployment of the operators.
	Therefore, the UAVs associated with different BSs may interfere with each other in uplink transmission as shown in Fig.~\ref{fig: sys mod}, which is referred to as the inter-cell interference.
	Since UAVs are likely to have LoS channels to multiple BSs owing to their high altitudes, the inter-cell interference may be severe in the cellular Internet of UAVs.
	%[SPACE]è¿æ®µå¯ä»¥å ä¸å¥è¯åé¡µæ°
\end{itemize}

%===================================
\subsection{Distributed Sense-and-Send Protocol}
%===================================

\begin{figure}[!t]
	\centering
\includegraphics[width=0.8\linewidth]{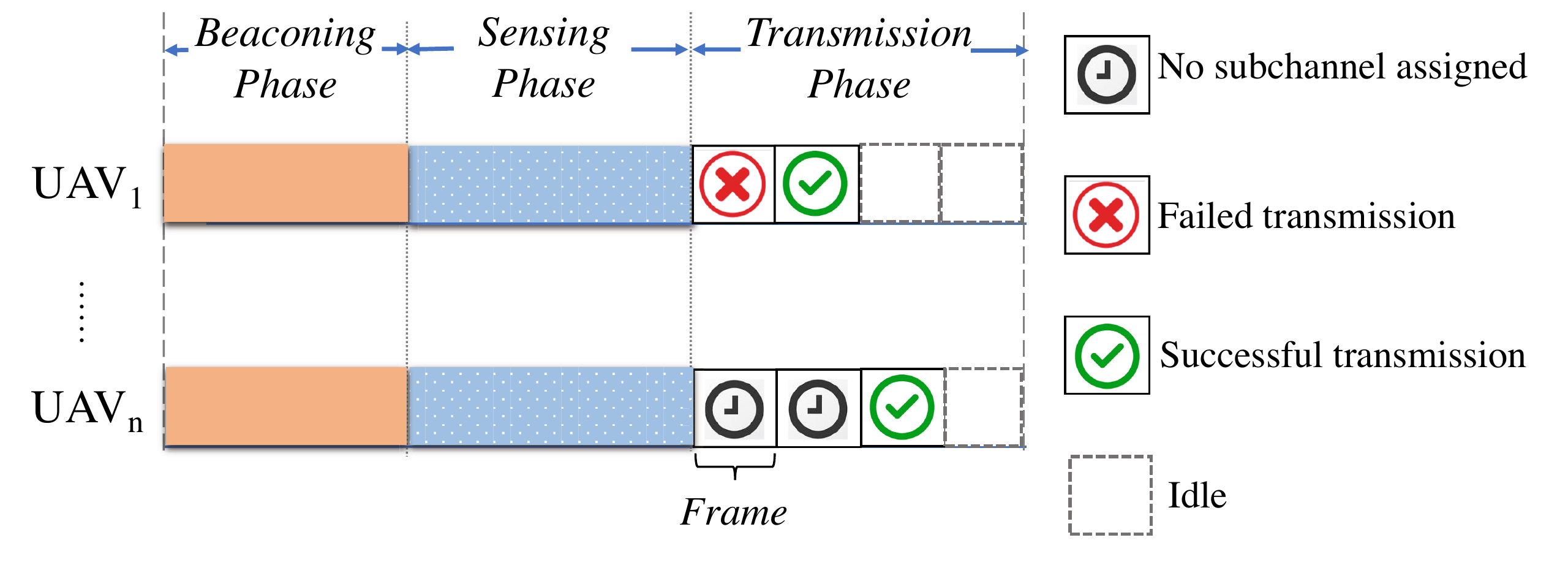}
		\vspace{-1.em}
	\caption{Sense-and-send cycle.} \label{fig: protocol}
		\vspace{-1em}
\end{figure}

%TODO è¿ä¸ªå¾æ¯éè¦ä¿®æ¹ç
%èèå é¤ç¬¬äºä¸ªcycleï¼è¯¦ç»ç»åºbeaconing phaseä¹ç±»çæ¹æ³ã
To coordinate multiple UAVs to execute the sensing tasks in a distributed manner, we propose the following distributed sense-and-send protocol based on~\cite{JHL-submitted}.
In this protocol, UAVs perform sensing and transmission in a synchronized iterative manner in the unit of sense-and-send cycle, or \emph{cycle} in short.
In each cycle, UAVs need to sense the targets and transmit the sensory data to the BS immediately.
As shown in Fig.~\ref{fig: protocol}, a cycle contains three phases: the \emph{beaconing phase}, the \emph{sensing phase} and the \emph{transmission phase}, which are explained as follows.

\subsubsection{Beaconing Phase}
\blue{At the beginning of the beaconing phase, each BS first broadcasts a beaconing frame on the wireless control channel, which contains the identity of the BS.
To synchronize the UAVs and the BSs, the synchronization signals adopted by cellular communications can be used in the beaconing frames~\cite{Fotouhi2019Survey}.
After receiving the beaconing frames, all the UAVs are synchronously informed that a new cycle has begun.
Then, the UAVs send back their state information to their associated BSs in the last cycle on the control channels, which includes their locations and the channel conditions to the BSs.
The BSs will exchange the state information of the UAVs with each other and then broadcast it on the wireless control channel, which is received by the UAVs.
By this means, the BSs can obtain necessary information from the UAVs to perform subchannel allocation.}

\blue{
Besides, based on the information provided by the BSs, the UAVs can make decisions on its sensing and transmission in a distributed manner, including their user associations, trajectories, and transmit power levels.
This decision-making process takes place at the end of the beaconing phase.
Moreover, as the duration of a cycle is pre-defined, the UAVs know when to expect the next beaconing frames from the BSs, and thus the synchronization will not be lost.}
%æåºä¸ä¸æ äººæºå¨beacon phase çæ«å°¾è¿è¡decising makingã

%-----------------------------------
\subsubsection{Sensing Phase}
%-----------------------------------
In the sensing phase, UAVs sense the targets of the tasks and collect sensory data.
We assume that the UAVs cannot determine whether the sensing was successful or not based on their sensory data, due to their limited onboard processing abilities.
Therefore, the UAVs need to send their sensory data to the BSs, and the BSs will decide whether the sensory data are valid or not.

%------------------------------------------
\subsubsection{Transmission Phase}
%------------------------------------------
In order to synchronize the transmissions of UAVs, the transmission phase of each cycle is further divided into \emph{frames}, which serve as the basic time unit for subchannel allocation.
The UAVs transmit their sensory data to the BSs through allocated subchannels in each frame. 
%If two UAVs are allocated with different subchannels, there will be no interference between them since the subchannels are orthogonal.
%However, when two UAVs from the same (or different) cell attempt to transmit in the same subchannel, they will suffer from intra-cell or (inter-cell) interference.
As shown in~Fig.~\ref{fig: protocol}, there are four possible situations that can occur to the UAVs in each frame of the transmission phase:
\begin{itemize}
\item \textbf{No subchannel assigned:} The UAV is allocated no subchannel to, and it needs to wait for the next frame.
\item \textbf{Failed transmission:} The UAV is allocated a subchannel to by the BS. However, the transmission has failed due to the low received signal power or the interference from other UAVs' transmission. Therefore, the UAV needs to transmit again in the following frames.
\item \textbf{Successful transmission:} The UAV is allocated a subchannel to, and it transmits the sensory data to the BS successfully.
\item \textbf{Idle:} The UAV keeps idle and will not transmit as it has already sent the sensory data to the BS successfully.
\end{itemize}

Since the spectrum resources are scarce, the subchannels of a BS may not be sufficient to support all the associated UAVs to transmit their sensory data in the same time.
To deal with this problem, the BSs need to adopt subchannel allocation mechanisms to allocate the limited number of subchannels to the UAVs efficiently.
An example of the subchannel allocation mechanisms is to allocate the subchannels to the UAVs that have the highest probabilities of successful transmission, which is adopted in~\cite{JHL-submitted}.
Besides, reinforcement learning can also be applied for the subchannel allocation, which will be discussed in Section~\ref{sec: application of RLs}.
%%TODO è¿ä¸é¢ä¸¤å¥æç¹åºè¯ï¼èèå æ
%Besides, it is also worth noticing that in a cycle, the number of UAVs requiring the subchannels decrease as frames passes by. 
%This is because UAVs who succeed in sending sensory data  keep idle in the rest of frames. 

%=============================================
\section{Reinforcement Learning for the Cellular Internet of UAVs}
%=============================================
\label{sec: RL and app.}
To better understand the applicability of reinforcement learning in the cellular Internet of UAVs, we start with introducing the basics of reinforcement learning.
We then categorize reinforcement learning approaches into four types, i.e. multi-bandit learning, Q-learning, actor-critic learning, and deep reinforcement learning, and give a brief introduction respectively.
Finally, we discuss possible applications of reinforcement learning to solve key problems in the cellular Internet of UAVs.

%-------------------------------------------------------
\subsection{Basics of Reinforcement Learning} 
%-------------------------------------------------------
%Simple Introduction of RL
Reinforcement learning is a learning process in which agents make decisions sequentially, observe the results, and then automatically adjust their policies for their objectives~\cite{E-2014}.
%TODO work here! ä»è¿ä»¥åé½æ¯åªç¨äºgrammarlyæ¥æ¥è¯­æ³éè¯¯çã
To be specific, reinforcement learning is based on the Markov decision process~(MDP), which consists of an environment, and some agents.
At each time step, the environment is at a certain state, and each agent selects a certain action according to its policy.
Then, the environment transits into a new state which is determined by its previous state and the actions of the agents.
 A reward is generated for each agent, which quantifies how well the objective of the agent is achieved.
Since the state transition is suitable to model the dynamics of the environment, the MDP is widely applied to model sequential decision-making problems.

%Long version
Besides, in contrast to centralized scheduling approaches and supervised machine learning, reinforcement learning does not rely on accurate prior knowledge of the environment or historical labeled data.
%Short version
%Besides, in contrast to centralized scheduling approaches and supervised machine learning, reinforcement learning does not rely on accurate prior knowledge on the environment or historical labeled data.
Instead, in reinforcement learning, the agents can automatically learn from the environment and their own past experiences to improve their policies.
This characteristic makes reinforcement learning suitable to be applied in the cellular Internet of UAVs, in which the UAVs face a rather dynamic and complex environment.
%Category of RL
Generally, reinforcement learning can be categorized into four types: multi-armed bandit learning, Q-learning, actor-critic learning, and deep reinforcement learning.
In the following, we will elaborate on these four reinforcement learning approaches.

\begin{figure}[!t]
	\centering
	\includegraphics[width=1 \linewidth]{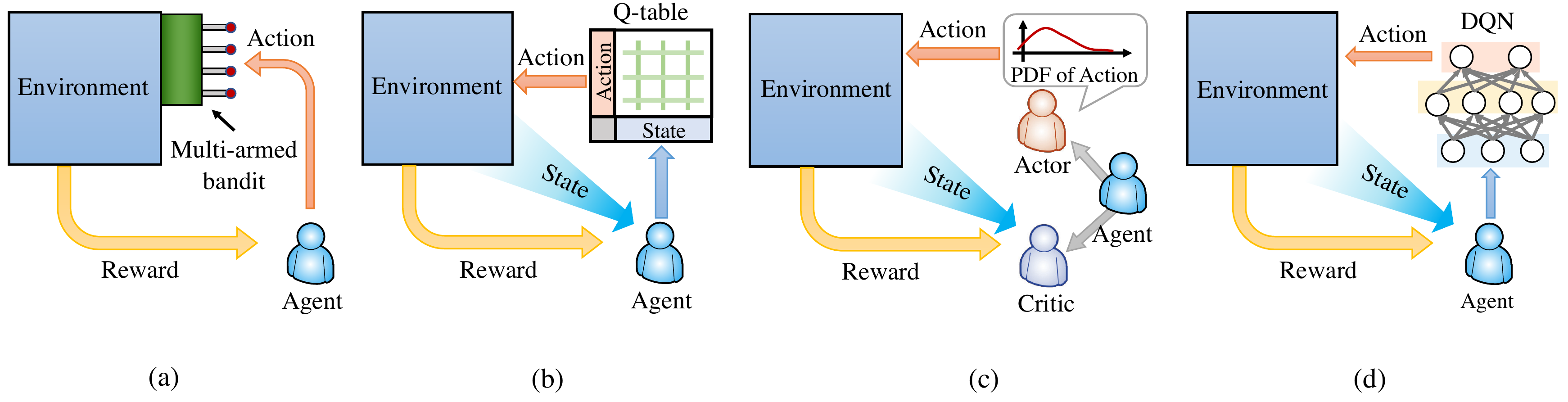}
	\vspace{-3em}
	\caption{Illustrations of~(a) multi-armed bandit learning,~(b) Q-learning,~(c) actor-critic learning, and~(d) deep reinforcement learning.} \label{fig: illustration}
		\vspace{-1em}
\end{figure}

%---------------------------------------------------
\subsubsection{Multi-armed Bandit Learning}
%---------------------------------------------------
As shown in Fig.~\ref{fig: illustration}~(a), in multi-armed bandit learning, the agent selects actions without recognizing the state of the environment.
After each time step, a reward is received by the agent, which is relevant to the action performed in the time step.
Based on the received rewards for each actions, the agent keeps track of the potential rewards associated with the actions.
The agent updates the potential reward estimations by a linear combination of the previous values in the table and the latest received rewards.
Since the agent aims to maximize the total rewards, it needs to select the current best action with the highest estimated reward.
Nevertheless, as the agent needs to explore the potential reward associated with each action in order to chose the best action, there exists a tradeoff between the \emph{exploration} and the \emph{exploitation}.
To be specific, the agent needs to decide to take the current best action or search for a better action.

Since the state of the environment has not been considered, the multi-armed bandit learning is inefficient to deal with fast-changing environments.
However, this disadvantage can be compensated by its very low complexity in implementation due to its low memory and computation requirements.
\blue{To be specific, suppose that the maximum number of available actions of the agent is $N$, and the agent needs to store the $N$ potential rewards associated with the actions.
At each time step, the agent needs to find the current best action and then updates the potential reward associated with the selected action, which can be represented by a linear combination of the previous estimation and the latest received reward.
Therefore, the computational complexity in each time step is $\mathcal O(N)$.
In summary, the multi-armed bandit learning is of very low memory requirement and computational complexity.}

%--------------------------------
\subsubsection{Q-learning}
%--------------------------------
As shown in Fig.~\ref{fig: illustration}~(b), the agent in Q-learning selects actions based on the \emph{Q-table}. 
To be specific, the value in Q-table, i.e. Q-value, for each state-action pair represents the estimated total rewards for the agent after executing the action at the state, under its current policy.
The agent adopts the policy to select the action which has the largest Q-value at each state.
To train the Q-table for more accurate estimation, after each time step, the agent updates its Q-table based on the observed reward.
Based on the updated Q-table, the policy of the agent also updates.
Therefore, the Q-value and the policy of the agent update iteratively, and eventually the policy will converge to the optimal policy of the agent~\cite{Sutton2018Reinforcement}.

\blue{To implement Q-learning, the agent who has $N$ available actions and $M$ states needs to store a Q-table involving $N\times M$ elements.
Besides, in each time step, the computations of an agent are similar to those of the multi-armed bandit learning, which indicates the computational complexity to be $\mathcal O(N)$.
Therefore, the memory requirement and computational complexity of the Q-learning is low.}

%--------------------------------------------
\subsubsection{Actor-critic Learning}
%--------------------------------------------
As shown in Fig.~\ref{fig: illustration}~(c), the agent in actor-critic learning are logically split into two roles, i.e. a \emph{critic} and an \emph{actor}.
The actor represents the action selection policy, which is a probability distribution function~(PDF) over the action space.
The critic observes the states and rewards from the environment and evaluates the state value, i.e. the expected total reward that will be received in the future passing through the state, which can be considered as a state value function.
The critic is used to improve the efficiency and stability for the training of the actor in term of optimal action selection.
Specifically, after each time step, the critic updates the state value based on the observed reward.
Then, the actor updates its policy for the previous state towards the direction which maximizes the performance.
%TODO è¿å¥è¯å¶å®è®©äººçä¸æ
Compared with other reinforcement learning algorithms, actor-critic learning does not search the action space for the action with the highest expected reward.
Instead, the action is selected randomly following the PDF provided by the actor.
Therefore, the complexity of action selection does not grow with the size of the action space.
This characteristic makes actor-critic learning more efficient in handling large or continuous action spaces compared to the other reinforcement learning approaches.

\blue{To implement actor-critic learning, the agent needs to store the action selection policy and the state value function.
    The action selection policy and state value function are both parameterized, and thus the amount of memory to store them is determined by the number of parameters quantifying them.
    In each time step, the agent needs to select an action according to the policy and compute the gradients of action selection policy and state value function with respect to the parameters to improves the actor and critic.
    In general, the action selection policy and state value function are the combinations of basic functional elements weighted by the parameters, e.g., linear, cosine, and exponent functions, and the number of parameters is small.
    Therefore, both the memory requirement and computational complexity of the actor-critic learning are low.}

%------------------------------------------------------
\subsubsection{Deep Reinforcement Learning}
%------------------------------------------------------
In deep reinforcement learning, deep neural networks are utilized to handle high-dimensional state spaces \cite{Arulkumaran2017Deep}.
% deep neural networks can automatically find the compact low-dimensional representations of high-dimensional states.
%In this approach, deep neural networks serve as components of agents.
As shown in Fig.~\ref{fig: illustration}~(d), at each time step, the agent inputs the feature vector representing the current state into the deep neural network, which can estimate the Q-value for each action, and thus is referred to as the deep Q-network~(DQN).
The agent then selects the action with the largest estimated Q-value, and stores the experience including the state transition and reward into a replay buffer, which is used to train the DQN to estimate Q-values more accurately.

\blue{To implement deep reinforcement learning, the agent needs to store the DQN and a replay buffer which stores its previous experiences.
Generally, to obtain better results, the sizes of the DQN and the replay buffer need to be large, which results in a high memory requirement.
The training of the DQN requires the gradients of estimated Q-values with respect to the parameters of the DQN.
This leads to a high computation complexity given a large DQN.
Therefore, deep reinforcement learning requires more memory and higher computational complexity than other reinforcement learning approaches.
Nevertheless, to handle the high requirements, the DQN can be trained in an offline manner.
For example, an agent can upload its experiences to a server.
The server trains the DQN according to the experiences and returns the DQN to the agent periodically.}

%===============================================
\subsection{Applications in the Cellular Internet of UAVs}
%===============================================
\label{sec: application of RLs}
\blue{Due to the rapid development of UAVs, UAVs can have enough onboard computation and memory capacities to perform reinforcement learning approaches, either by having in-built circuits or carrying additional computing devices.
This allows the implementations of reinforcement learning approaches to solve key problems in the cellular Internet of UAVs.}
In the following part, we discuss possible implementations of different types of reinforcement learning to solve the trajectory control and resource management problems in the cellular Internet of UAVs, including user association, power management, the subchannel allocation.

%**è¿éåºç¨çé»è¾æç®è¿ä¹åï¼ç¬¬ä¸æ®µæè¿°é®é¢ï¼ç¬¬äºæ®µæè¿°å»ºæ¨¡ï¼ç¬¬ä¸æ®µæè¿°åéçåå ã

%TODO æ é¢æ ¼å¼åé¢é½æ¹ï¼ä¹çä¸ä¸introéé¢ææ²¡æéè¦æ¹ç
%--------------------------------------------------------------------------
\subsubsection{Multi-armed Bandit Learning for User Association}
%--------------------------------------------------------------------------

In the cellular Internet of UAVs, each BS is equipped with $K$ subchannels to support at most $K$ UAVs to transmit simultaneously, and each UAV needs to choose one BS to associate with, which is called as the user association. 
According to the proposed protocol in Section~\ref{sec: Scenario}, the probabilities of successful transmission for UAVs decrease with the number of UAVs associated with the same BS increasing, due to the aggravated competition for the subchannels.
Besides, due to the shadowing in the air-to-ground channels, the channel conditions from a UAV to different BSs are usually varying.
For example, when the channel between the UAV and the BS has an LoS component, its pathloss is much lower than when no LoS component exists due to obstruction~\cite{3GPP-2017}.
For the above reasons, the user association is a challenging problem in the cellular Internet of UAVs.

Since multi-armed bandit learning does not rely on prior information to predict the channel conditions and the numbers of associated UAVs of the BSs, it is suitable for the user association problem.
At the beginning of each beaconing phase, each UAV as agent decides which BS to associate with.
As the objective of the UAV is to maximize the successful transmission probability, the reward is $1$ if the sensory data is successfully transmitted; otherwise, the reward is $0$.
Each UAV optimizes its action selection policy by estimating the expected reward for each action and selecting the action with the highest current expected reward.
Besides, the UAV also maintains an exploration probability, with which it selects an action randomly, in order to keep searching for a better action.

%---------------------------------------------------------
\subsubsection{Q-learning for Trajectory Control} 
%---------------------------------------------------------
\blue{When the UAV executes a sensing task, the successful sensing probability will be higher if it gets closer to the sensing target. 
On the other hand, the UAV will have a higher probability to transmit the sensory data to the BS successfully, if it approaches the BS.}
Therefore, sensing and transmission are coupled by its trajectory.
However, since accurate sensing and transmission models are hard to obtain in the complex and dynamic environment of cellular Internet of UAVs,  the trajectory control is also challenging.

As Q-learning does not require models of the coupled sensing and transmission, it is suitable for the trajectory control problem.
To apply Q-learning, the flying space can be abstracted into a finite set of discrete spatial points, and the trajectory can be considered as a path through these spatial points. 
The state of each UAV is its location, and the action is its trajectory in each cycle.
Since the objective of each UAV is to send valid sensory data to the BS, the reward is $1$ if valid sensory data received successfully by the BS; otherwise, the reward is $0$.
%----------------------------------------------------------------------
\subsubsection{Actor-critic Learning for Power Management} 
%----------------------------------------------------------------------
In the cellular Internet of UAVs, the UAVs need to raise their transmit power in order to satisfy the QoS constraint.
However, the onboard batteries of UAVs are generally highly limited; therefore, high transmit power results in the batteries draining quickly.
In order to maximize the total number of successful transmissions before the battery run out, it is crucial for each UAV to adopt efficient approaches for transmit power management.

Since the transmit power level can take values from continuous space, it is suitable to apply actor-critic learning for the power management problem.
To be specific, in each cycle, the pathloss of the uplink subchannel and remaining battery capacity can be jointly considered as the state.
At the end of the beaconing phase, the actor of UAV selects a transmit power level based on the PDF over actions according to the current state.
After the cycle, the reward to the UAV is $1$ if the transmission was successful; otherwise, the reward is $0$.
The experience, consisting of the state transition and the reward, is used to train the critic to estimate the value of state more accurately.
Then, the actor updates itself to select a better action by learning from the experience and the critic's evaluation.

%--------------------------------------------------------------------------------------
\subsubsection{Deep Reinforcement Learning for Subchannel Allocation}
%--------------------------------------------------------------------------------------
\label{sec: subchannel allocation-DRL}
In the cellular Internet of UAVs, the uplink transmissions of UAVs may suffer severe inter-cell interference, due to the low pathloss of LoS channels between each UAV and multiple BSs.
To handle this problem, the BSs that share the same frequency band need to perform subchannel allocation jointly, in order to alleviate the inter-cell interference and improve the probability of successful transmission.

However, in the subchannel allocation, the channel states between multiple BSs and their associated UAVs need to be taken into consideration.
Besides, since the UAVs who are idle in the frame do not transmit, the transmission situations of UAVs should also be considered.
Therefore, the subchannel allocation problem has a complex and high-dimensional state space.
As deep reinforcement learning is able to solve the optimal policies for agents facing high-dimensional state space, it is suitable for the subchannel allocation problem in the cellular Internet of UAVs.

In this case, the BSs that share a certain frequency band can be considered as the agent, and the action of the BSs is the subchannel allocation for the UAVs.
In each frame in the transmission phase, the state composes of the following elements:
\begin{itemize}
\item the pathlosses for the channels between the involved BSs and their associated UAVs.
\item the indicators for whether the UAVs are idle in the frame.
\end{itemize}

At each frame in the transmission phase, the BSs form a feature vector representing the current state and input it into the DQN, which returns the Q-value for each possible subchannel allocation of the UAVs. 
Then, the BSs select the subchannel allocation with the largest Q-value as its action.
After the cycle, the reward given to the agent can be designed as the number of successful transmissions.
The rewards along with the transition among states are stored at the replay buffer of the UAV, which is then used to train the DQN to estimate the Q-values for state-action pairs more accurately.

In Table \ref{Summary}, we summarize the different types of reinforcement learning approaches with their characteristics and their applications in the cellular Internet of UAVs.

\begin{table}
\scriptsize
	\centering
	\caption{Reinforcement learning for the cellular Internet of UAVs} \label{Summary}
	\vspace{-1em}
	\begin{tabular}{| p{3.6cm}<{\centering}| p{7cm}<{\centering}|p{3.0cm}<{\centering} |}
\Xhline{1.pt}
		\thickvlines \textbf{Reinforcement learning} & \textbf{Characteristics} & \textbf{Applications} \\
		\Xhline{1.pt}\hline
		 \makecell[l]{Multi-armed bandit learning} &  \makecell[l]{Very low complexity;\\Consider no environment state;} &\makecell[l]{User association} \\
		\hline \makecell[l]{Q-learning} &  \makecell[l]{Low complexity;\\Consider state transitions;\\Inefficient to deal with large action spaces.} & \makecell[l]{Trajectory control}\\
		\hline \makecell[l]{Actor-critic learning} &  \makecell[l]{Low complexity;\\Consider state transitions;\\Capability to handle large or continuous action spaces.} & \makecell[l]{Power Management}\\
		\hline \makecell[l]{Deep reinforcement learning~} &  \makecell[l]{ High complexity;\\Consider state transitions;\\Capabaility to handle large state spaces.} & \makecell[l]{Subchannel allocation}\\
		\Xhline{1.pt}
	\end{tabular}
	\vspace{-1em}
\end{table}

%%%%%%%%%%%%%%%%%%%%%%%%%%%%%%%%%%%%%%
\section{Reinforcement Learning Example: Q-Learning for Trajectory Control}
%%%%%%%%%%%%%%%%%%%%%%%%%%%%%%%%%%%%%%
\label{sec: Q-example}
As an illustrative example, in this section, we introduce how to apply the reinforcement learning approach to solve the trajectory control problem in the cellular Internet of UAVs. 
%First, we describe the trajectory design problem in UAV sensing network.
%Then, we propose the reinforcement learning framework to formulate the trajectory design problem.  
%Within the framework, we adopt multi-agent Q-learning algorithm to solve the problem, and further enhance it by action-set reduction and model-based reward representation. 
%=======================
\subsection{Problem Description}
%=======================
\label{sec: trajectory control problem}
We consider a cellular Internet of UAVs which consists of two BSs and three UAVs associated with each BS. 
These two BSs are assumed to have two subchannels to support uplink transmissions of UAVs, and the frequency bands utilized by different BSs are different. 
Moreover, the UAVs are flying in a cylindrical space centered at the BS, with the minimum and maximum flying altitudes constraints.

To tackle the trajectory efficiently, we divide the space into a finite set of discrete spatial points, which are arranged in a square lattice pattern.
The trajectory control problem can be reformulated as each UAV selecting an adjacent spatial point of the current spatial point in each cycle to maximize the total rewards, i.e. the number of valid sensory data successfully sent to the associated BS. 
The distance from the current point to the next one should be less than UAVs' maximum flying distance in a cycle.
Note that for each UAV, the valid sensory data received by the BS in future worths less than that received currently, which means the UAVs have discount values on their future rewards. 

 %=====================
\subsection{Algorithm Design}
 %=====================
\label{sec: RL algorithm}
To solve the trajectory control problem, we utilize the Q-learning algorithm where the UAVs are considered as agents, the locations of UAVs is the state, and the cycle is the time unit.
The available action set for a UAV consists of all the possible linear flying trajectories to the feasible spatial points whose distance is less than the maximum flying distance of a UAV in a cycle.
The state transition is the function which maps the current locations and actions of UAVs to the locations of the UAVs at the beginning of the next cycle.
Finally, the reward is $1$ if the BS receives the valid sensory data successfully from the UAV in the cycle; otherwise, the reward is $0$.

To be specific, the Q-learning algorithm for the trajectory control problem can be briefly described as follows:~(1) At the beginning of a cycle, each UAV chooses the action which has the maximum Q-value in the current state;~(2) The UAV performs the selected action in the cycle;~(3) At the beginning of the next state, the UAV observes the transition state and is informed whether valid sensory data is received by the BS;~(4) The UAV updates the state-action value in the previous state, and then select an action for the new state.

Moreover, to improve the efficiency of Q-learning for the trajectory control, we propose an enhanced multi-UAV Q-learning algorithm in~\cite{JHL-submitted}, which is based on the following observations.
First, since all the UAVs determine trajectories at the same time, each UAV needs to consider other UAVs' action selections when selecting its own.
This can be achieved by that each UAV keeps a record on other UAVs' action selection statistics at each state~\cite{claus1998the}.

Secondly, it can be observed that if the UAV does not locate near the vertical plane passing through the BS and the target, both the successful sensing and transmission probabilities will decrease. 
Therefore, the available action set of each UAV can be reduced to the actions towards or on the BS-target plane.

Thirdly, in the traditional Q-learning algorithms, agents update their values using the observed reward in the last cycle. 
However, in such a manner, the estimated Q-values converge slowly, and the performance of the algorithms are likely to be poor.
%[SPACE] long version
Therefore, in the proposed algorithm, UAVs update their Q-functions based on the probability of successful valid sensory data transmission, which can be calculated by the algorithm proposed in~\cite{JHL-submitted}.
%The UAVs can calculate the probability of successful valid sensory data transmission by using the algorithm proposed in~\cite{JHL-submitted}. 
%short version: ... which can be calculated ...

%, in which the available action set is reduced adopts the following methods, i.e., available action set reduction and model-based reward representation.
%
%\begin{figure}[!t]
%	\centering
%	\includegraphics[width=0.63\linewidth]{figures/simulation/simul_figure_2.eps}
%%		\vspace{-1.em}
%	\caption{UAVs' average per cycle reward versus the number of training cycles.}
%	\label{fig: performance of proposed UAV Q-learning}
%		\vspace{-1em}
%\end{figure}

 \begin{figure}[!t]
\centering
    \subfigure[]{
        \begin{minipage}[t]{0.5\linewidth}
            \includegraphics[width=1\linewidth]{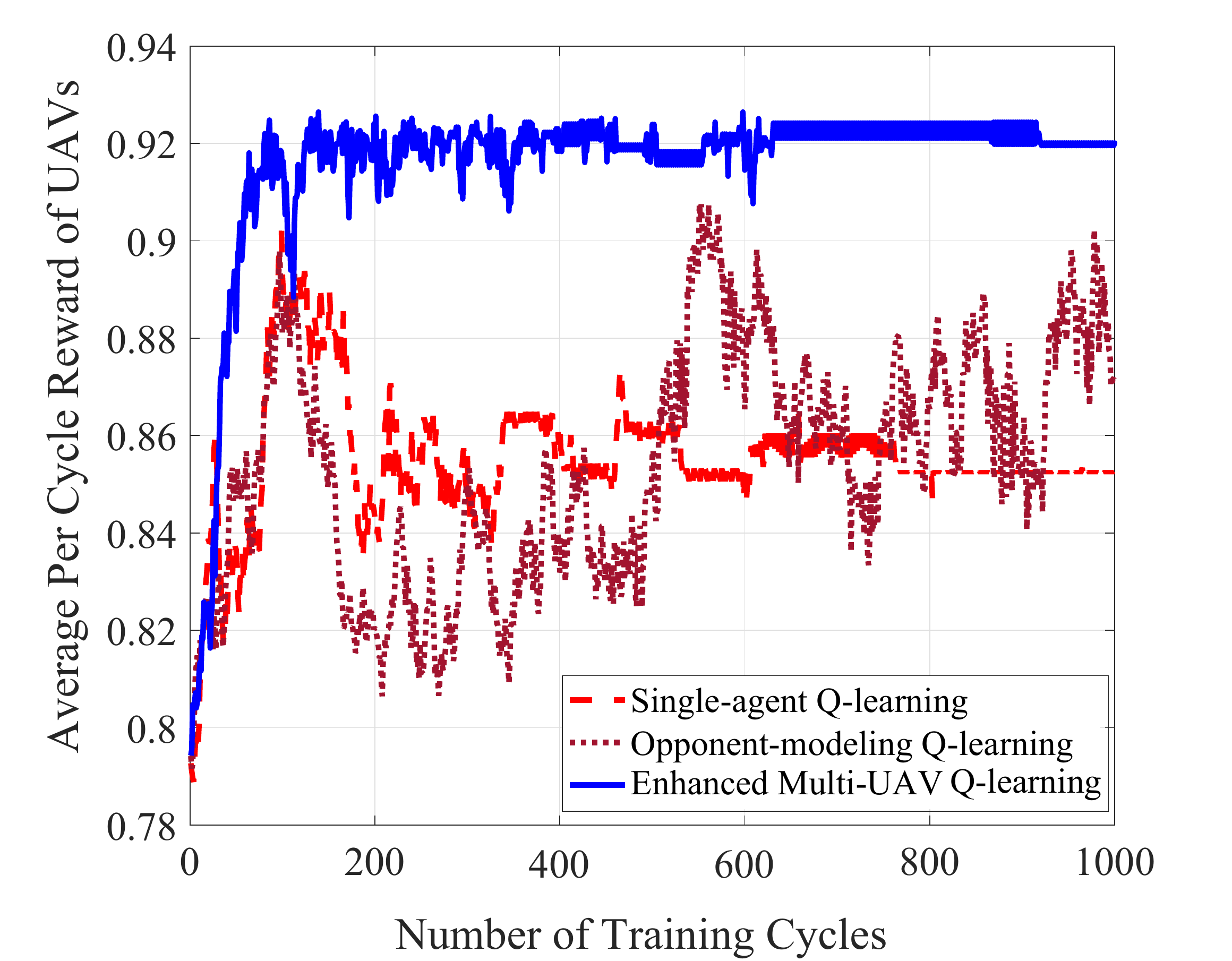}
        \end{minipage}
        }\subfigure[]{
        \begin{minipage}[t]{0.5\linewidth}
            \includegraphics[width=0.95\linewidth]{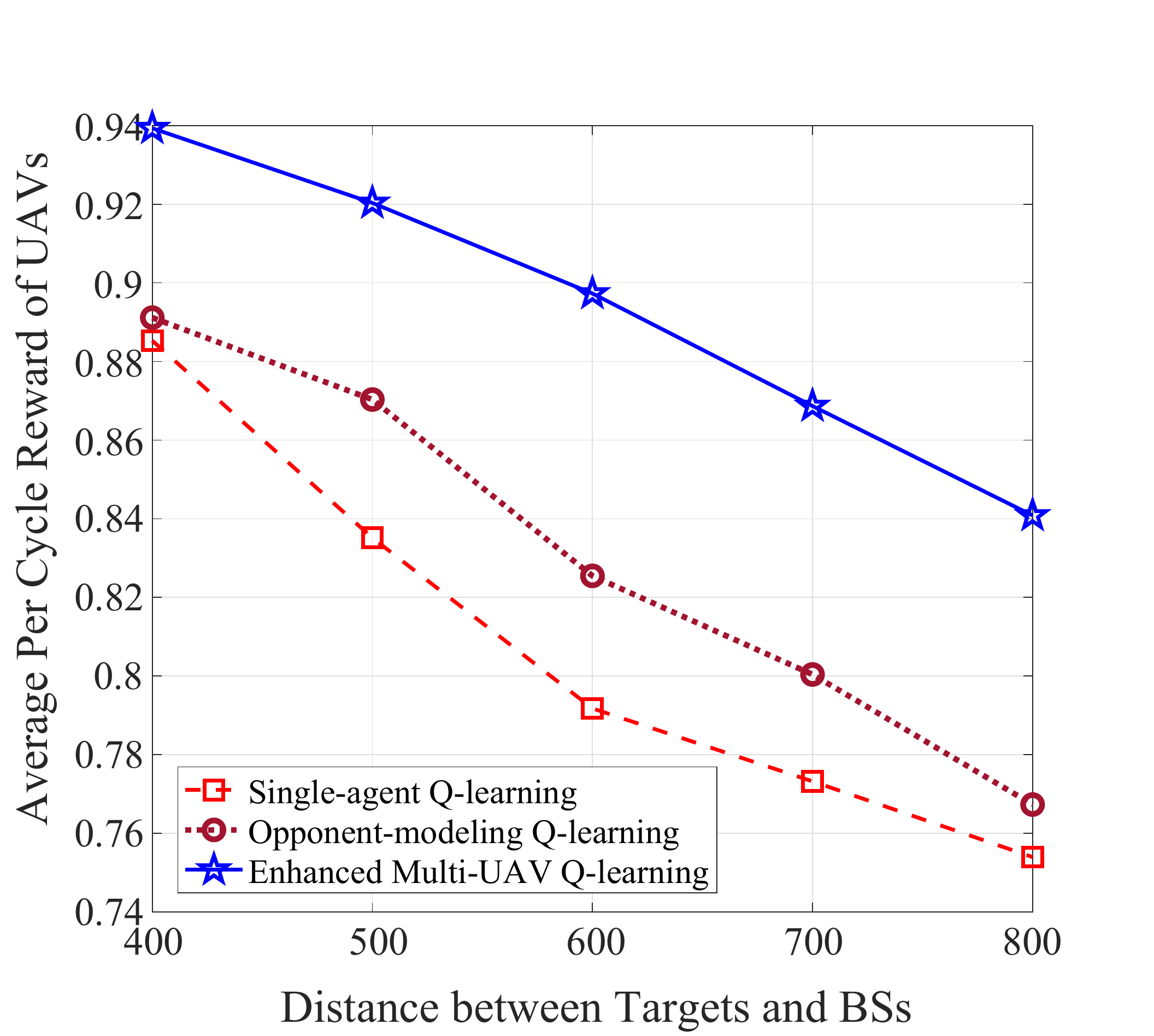}
        \end{minipage}
        }
        \vspace{-1.5em}
	\caption{UAVs' average per cycle reward versus~(a) number of training cycles;~(b) distance between the targets and the BSs.} \label{fig: performance of proposed UAV Q-learning}
	\vspace{-1em}
\end{figure}

To evaluate the performance of the proposed enhanced multi-UAV Q-learning algorithm, we compare it to the traditional single-agent and opponent modeling Q-learning algorithms~\cite{claus1998the}. 
Fig.~\ref{fig: performance of proposed UAV Q-learning}~(a) shows the average per cycle reward of the UAVs versus the number of training cycles.
It can be seen that compared to the traditional Q-learning algorithms, the proposed algorithm converges faster and to a higher reward, i.e. the number of valid sensory data received by the BSs, in each cycle.
Besides, Fig.~\ref{fig: performance of proposed UAV Q-learning}~(b) shows the average per cycle reward of the UAVs versus the distance between the targets and the BSs.
It indicates that the probability of successful valid data transmission decreases with the distance between the targets and the BSs.
Nevertheless, the decrement in the proposed algorithm is less than those in the other algorithms. 
This shows that the proposed algorithm is more robust to the variance of the targetsâ locations.

%%%%%%%%%%%%%%%%%%%%%
\section{Conclusions and Future Outlook}
%%%%%%%%%%%%%%%%%%%%%
\label{sec: Conclusion}

%TODO æ¶æè¦ç»ä¸ç¨ç°å¨å®ææ¶

In this article, reinforcement learning has been introduced as a distributed approach to solve key problems in the cellular Internet of UAVs. 
We have introduced the cellular Internet of UAVs and then proposed a distributed sense-and-send protocol for the coordination of multiple UAVs to execute sensing tasks.
Following that, we have introduced the basics of reinforcement learning and specified the four categories of reinforcement learning approaches.
\blue{We have also discussed the potential applications of reinforcement learning approaches to tackle the trajectory control and resource management problems in the cellular Internet of UAVs.
To provide an example, we have elaborated on using the enhanced multi-UAV Q-learning algorithm to solve the trajectory control problem.}

As a powerful approach for UAVs to automatically learn optimal decision-making policies from past experiences, reinforcement learning is promising for the cellular Internet of UAVs.
Nevertheless, there are still many open issues in this field, which may drive future inventions and researches. 
Several potential research directions are listed below as examples.

%=================================
\textbf{Cooperative Cellular Internet of UAVs:}
%=================================
When the targets are far away from the coverage of the BSs, the UAVs may need cooperation to execute the tasks.
%ä¸sensingçæ¶åä½ä¸ºrelay
To be specific, a UAV can choose not to sense any targets, but to work as a relay which helps another UAV transmit sensory data to the BSs~\cite{Zhang2018Joint}.
In this case, the UAVs need to choose their roles in each cycle, with the objective to maximize the number of valid sensory data received by the BS.
To tackle this problem, Q-learning can be applied.
Specifically, the state can be the locations of the UAVs and the BSs, and the actions of the UAVs are their decisions on whether to sense or to relay.
Moreover, to optimize the overall performance in terms of transmitting valid sensory data, the rewards can be defined as the number of valid sensory data received by the BS in each cycle.

\textbf{Cognitive Cellular Internet of UAVs:} 
%TODO add more words
In some sensing tasks, e.g., live streaming, the transmission of a large amount of sensory data generated by UAVs may pose a significant burden on cellular networks. 
To guarantee the quality of service for traditional cellular users while improving the quality of data transmission for UAVs, cognitive radio can be utilized to enable UAVs to opportunistically access the channels which are originally occupied by the cellular users. 
Under such a setting, cellular users and UAVs serve as the primary and secondary users, respectively.
This channel access problem can be solved by deep reinforcement learning \cite{Wang2018Deep}, where UAVs are the agents while the combination of the previous observations on subchannels is regarded as the state.
To avoid interfering with primary users, the rewards of UAVs can be designed as a weighted sum of the successful transmission probability and the interference incurred to the primary users. 

\textbf{Millimeter-wave Cellular Internet of UAV:} 
In the cellular Internet of UAVs, sensory data need to be transmitted to the BSs timely, which may require high data rates between the UAVs and the BSs.
Therefore, it is promising to apply millimeter wave~(mmWave) communication in the cellular Internet of UAVs, which can provide abundant frequency spectrum resource and alleviate the inter-cell interference due to its high attenuation rate.
To implement mmWave communication, beamforming is required at UAVs to steer strong Signal Noise Ratio~(SNR) at the BS, in which the UAVs need to search a large number of beam directions and finds the best angle.
To solve this problem, actor-critic learning can be adopted.
Specifically, the location of UAVs and BSs can be jointly considered as the state, and the beam directions can be considered as the UAVâs action space.
In each cycle, each UAV selects a beam direction according to the PDF generated by its actor part.
The rewards for UAVs can be designed as the received SNR level at the BSs in the cycle, in accordance with the objective to find the optimal beam direction.

%latency; mobility; synchronization; heterogeneous network
%\bibliographystyle{IEEEtran}%
%\bibliography{bibilio}
%
%

\end{document}